\documentclass{moriond}

\usepackage{amsmath,amssymb,amsfonts, bm,bbm,slashed}

\def\be{\begin{equation}}
\def\ee{\end{equation}}
\def\bea{\begin{eqnarray}}
\def\eea{\end{eqnarray}}

\newcommand{\Photo}{\includegraphics[height=35mm]{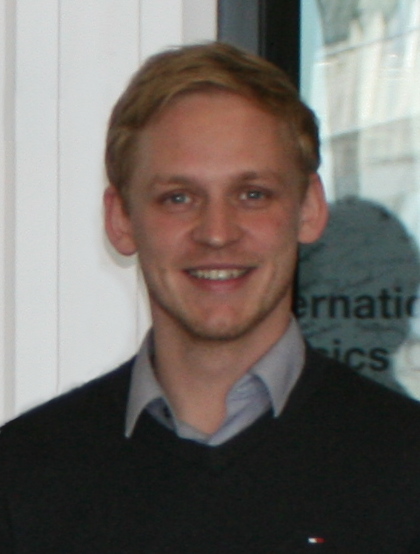}}

\begin{document}
\vspace*{4cm}
\title{Effective Field Theory after a New-Physics Discovery}

\author{Matthias K\"onig}

\address{Physik Institut, Universit\"at Z\"urich, CH-8057 Z\"urich, Switzerland}

\maketitle\abstracts{
After the discovery of a new resonance, its couplings to the Standard Model (SM) need to be described by the means of an effective theory, appropriately constructed to separate its mass scale from the mass scales associated with the SM decay products. At the example of a scalar resonance transforming as a singlet under the SM gauge group, we construct the operator basis for such a theory using the language of Soft-Collinear Effective Theory (SCET) and perform the resummation of the leading Sudakov logarithms.}

\section{Introduction}
To describe interactions between possible high-scale new-physics (NP) effects and the SM, one needs to account for the large scale gap between the NP scale and the scale of the observable under consideration. For example, it is well-known that large QCD logarithms $\alpha_s\log(q^2/\Lambda_\mathrm{NP}^2)$ can spoil the convergence of the perturbation expansion for hadronic low-energy observables, where $q^2\sim\mathcal O (\Lambda^2_\mathrm{QCD})$. By describing the process in terms of an effective field theory (EFT), these large logarithms can be resummed using renormalization group methods. For decays of a NP resonance, $q^2\sim \Lambda^2_\mathrm{NP}$ is of the high scale and one might suspect the absence of large logarithms. However, the scale hierarchy between $q^2$ and the masses in the final state introduces large Sudakov logarithms, which need to be resummed just as well.

The appropriate effective theory of a heavy-to-light transition is the Soft-Collinear Effective Theory~\cite{Bauer:2000yr,Bauer:2001ct,Bauer:2001yt,Beneke:2002ph}, in which the low-energy degrees of freedom are comprised of fields with low virtualities $k^2\sim 0$. In contrast to traditional EFTs, individual components of $k^\mu$ can still be large, as long as $k^\mu$ is light-like, and consequently operators can be non-local along the directions of these large momentum components. The interested reader is referred to the literature for more details on the construction and the resulting field-theory implications~\cite{Becher:2014oda}.

Here we show an excerpt of the operator basis of the recently developed SCET$_\mathrm{BSM}$, constructed for a singlet scalar resonance decaying to SM fields. Furthermore, we comment on the matching procedure to a concrete UV completion and the size of the resummation effects. The full operator basis along with the anomalous dimensions up to next-to-next-to-leading order in the EFT power-counting can be found in the original paper~\cite{Alte:2018nbn}, whereas the in-depth treatment of the UV completion was presented in a later work~\cite{Alte:2019iug}.

\section{Effective Lagrangian}
To construct the operator basis, one writes down all possible interactions between the resonance and the SM degrees of freedom allowed by the symmetries of the theory, just like one does in any EFT like for example the SMEFT~\cite{Weinberg:1979sa,Wilczek:1979hc,Buchmuller:1985jz,Leung:1984ni,Grzadkowski:2010es}. One could now envision supplementing the SMEFT operator basis by local operators coupling the scalar resonance $S$ to various SM singlet currents to describe the $S$ couplings to the SM fields~\cite{Gripaios:2016xuo,Franceschini:2016gxv}. Such a ``SMEFT+$S$'' would separate the scale of the resonance mass $m_S$ from the scale $m_\Psi$ of the UV completion that generates these couplings, but not the resonance mass from the SM mass scales $\mu_\mathrm{SM}\sim v$. Furthermore, there is no reason to believe that $m_\Psi \gg m_S$ should hold. In fact, it is very plausible that a new resonance is simply the first discovered particle out of a larger (undiscovered) NP sector. 

\begin{figure}
\centering
 \includegraphics[scale=.9]{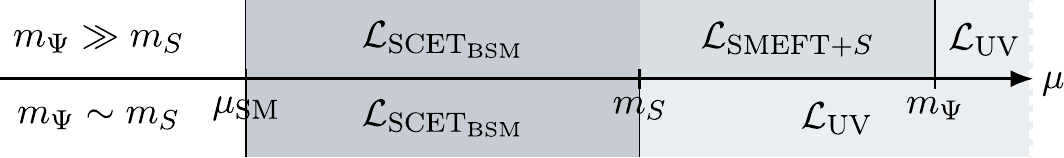}
 \caption{Validity regions and matching thresholds for the various effective theories depending on the scale hierarchy between the resonance mass $m_S$ and the scale of the additional NP sector $m_\Psi$.}\label{fig:scales}
\end{figure}

By constructing the operator basis in the language of SCET, one obtains an effective theory that can deal with both cases $m_\Psi \sim m_S$ and $m_\Psi \gg m_S$. In the latter case, one simply integrates out the UV completion at $m_\Psi$ and matches the ``SMEFT+$S$'' to the SCET at the matching scale $m_S$. If the sectors are at similar scales, $m_\psi \sim m_S$, one integrates out the full NP sector at $m_S$ and matches it to the SCET, as depicted in Fig.~\ref{fig:scales}.

In the SCET$_\mathrm{BSM}$, operators are organized in powers of the scale ratio $\lambda=\mu_\mathrm{SM}/m_S$. Instead of discussing the full basis, we will focus on three operators for brevity:
\begin{align}
 O_{AA} &= S\, g_{\mu\nu}^\perp \mathcal A_{n_1}^{\mu,a} \mathcal A_{n_2}^{\nu,a}\, &
 O_{\phi\phi}&= S\,\left(\Phi^\dagger_{n_1}\Phi_{n_2}+\Phi^\dagger_{n_2}\Phi_{n_1} \right) \,, &
 O^{ij}_{F_L \bar f_R} &= S\, \bar F^i_{L,n_1}\Phi_0 f^j_{R,n_2}\,,
\end{align}
where the first two are of $\mathcal{O}(\lambda^2)$ and the third one is of $\mathcal{O}(\lambda^3)$ in SCET power-counting. 
The subscripts $n_i$ denote the directions of large momentum in which the particle is moving, described by light-like reference vectors, satisfying~\footnote{The standard choice for a two-body decay would be $n_1 = (1,0,0,1)$, $n_2 = (1,0,0,-1)$, $\bar n_1 = n_2$, $\bar n_2 = n_1$.} $n^2=0$ and $\bar n\cdot n=2$. Each field operator can be displaced along this direction $\psi_n \equiv \psi_n(x+t\bar n)$. The Lagrangian is then a convolution over the parameters $t$ with the Wilson coefficients:
\begin{align}
 \mathcal L \sim \int dt_1\, dt_2\,C(t_1,t_2,\mu)\,S(x)\, \phi^\dagger_{n_1}(x+t_1\bar n_1)\phi_{n_2}(x+t_2\bar n_2)\,.
\end{align}
Field operators are dressed with Wilson lines, defined by:
\begin{align}
  W_{n}^{(A)}(x) = P\,\exp\left[i g_A\int_{-\infty}^0ds\,\bar n_i\cdot A_{n_i}(x+s\bar n_i)\right]\,,
\end{align}
to ensure gauge invariance. For scalars, fermions and gauge fields, the field operators are related to the SM fields via~\cite{Bauer:2002nz,Hill:2002vw}:
\begin{align}
  \Phi_n(x) &= W^\dagger_n (x)\phi(x)\,,&
  \Psi_n(x) &= \frac{\slashed n\slashed{\bar n}}{4}W_n^\dagger (x)\psi(x)\,,&
  \mathcal A^\mu_n(x) &= W^{(A)\dagger}_n(x)[iD^\mu_n W_n^{(A)}(x)]\,.
\end{align}
The field $\Phi_0$ denotes a Higgs doublet carrying no momentum, which will be replaced by the Higgs vacuum expectation value after electroweak symmetry breaking. The operator $O_{AA}$ generates the decays $S\to jj$, $\gamma\gamma$, $W^+W^-$ and $Z Z$ whereas the operator $O_{\phi\phi}$ is responsible for the di-Higgs decay as well as the decay into longitudinal electroweak bosons. The operators $O^{ij}_{F_L\bar f_R}$ generate the various difermion decays.

When performing the matching to the SCET$_\mathrm{BSM}$, one expands amplitudes in the mass ratio $\lambda$. If another NP scale $m_\Psi\sim m_S$ is present, one keeps the full dependence on it. As an example, the decay $S\to gg$, generated through a loop of vectorlike fermions of mass $m_\Psi$ leads to the Wilson coefficient
\begin{align}
 C_{GG}(q^2) = \frac{T_F}{\pi^2}\left[\left(\frac{4m_\Psi^2}{q^2}-1 \right)\arcsin^2\left( \sqrt{\frac{q^2}{4m_\Psi^2}}\right) - 1  \right]\,,
\end{align}
where the non-polynomial dependence of the coefficient on the momentum transfer $q^2=m_S^2$ is a consequence of the non-locality of the operator and a feature typical of SCET.

\section{Resummation of large logarithms}
The motivation for describing the decays $S\to \mathrm{SM}$ in the framework of the SCET$_\mathrm{BSM}$ is the systematic resummation of large (double) logarithms of the form $\alpha \log^2\lambda$. This is achieved by solving the renormalization group equations of the couplings in the effective theory. We will demonstrate this by showing the numerical coefficients $U_{i}(\mu_0, M)=C_i(\mu_0)/C_i(M)$, assuming a NP-scale $M = 2.5$~TeV. We will also show the ratio between the resummed and the fixed-order decay rates $R_i=\Gamma_i^\mathrm{res}/\Gamma_i^\mathrm{fo}$ for the modes $S\to\gamma\gamma$, $S\to jj$, $S\to t \bar t$ and $S\to h h$. Only numerical results will be shown here and none of the ingredients necessary to obtain them. They are detailed in the original papers and references therein~\cite{Alte:2018nbn,Alte:2019iug}, including analytical solutions to the RG equations governing the scaling behavior of the various Wilson coefficients.

Two operators contribute to the diphoton decay, $O_{BB}$ and $O_{WW}$. Their running is described by the coefficients $U_{WW}(m_W,M)=0.80\,e^{.23i}$, and $ U_{BB}(m_W,M)=1$. 
Neglecting an interaction of the form $S\phi^\dagger\phi$, the resummation leads to a suppression of the decay rate by a factor of:
\begin{align}
 R_{S\to \gamma\gamma}=|0.9 U_{WW}(m_W,M)+0.1|^2\approx 0.67\,,
\end{align}
which is a sizeable effect, originating solely from electroweak corrections.

The leading contribution to the decay of $S\to jj$ is given by the operator $O_{GG}$. The Wilson coefficient of this operator needs to be scale-evolved to $\mu_j$, which is an energy scale associated with the definition of the jets. Assuming $\mu_j = 100$~GeV, we find $U_{GG}\approx 0.38\,e^{0.98i}$. This leads to a suppression of the dijet rate by:
\begin{align}
  R_{S\to gg}=|U_{GG}(\mu_j,M)|^2\approx 0.15\,.
\end{align}
Neglecting resummation effects would thus vastly overestimate the decay rate due to large QCD corrections.

As an example of a difermion decay, the operator $C^{33}_{Q_L\bar u _R}$ generates the decay $S\to t\bar t$. The RG evolution from $M$ to the top-quark mass is given by $U_{q\bar q}(m_t,M)=0.90\,e^{0.31i}$. This is a relatively mild correction, leading only to a suppression of:
\begin{align}
  R_{S\to t\bar t}=|U_{q\bar q}(m_t,M)|^2\approx 0.81\,.
\end{align}

Finally, the di-Higgs decay of the $S$ is described by the operator $O_{\phi\phi}$. Solving its evolution equation yields the correction factor $U_{\phi\phi}(m_h,M)\approx 0.79\,e^{0.08i}$. Consequently, the di-Higgs decay rate is suppressed by:
\begin{align}
 R_{S\to hh}=|U_{\phi\phi}(m_h,M)|^2\approx 0.62\,,
\end{align}
which is again a large correction, despite the fact that the final state does not contain any color-charged particles.

\section{Conclusions}
We have developed an effective theory to describe the decays of a hypothetical scalar new-physics resonance into SM fields, the SCET$_\mathrm{BSM}$. Since the mass of the resonance injects large energies into the light final states, radiative corrections are bound to generate Sudakov logarithms of the form $\alpha \log^2(\mu^2_\mathrm{SM}/m_S^2)$, which can spoil the convergence of the perturbation expansion and need to be resummed to all orders. Since the light final state particles travel with large momenta, the appropriate effective theory is SCET, which can separate the scales $m_S$ and $\mu_\mathrm{SM}$.

The construction of the SCET$_\mathrm{BSM}$ does not make any assumptions about how the couplings between $S$ and the SM are generated in a UV-complete model, especially about the masses of a possible larger NP sector. Traditional EFT constructions, akin to the SMEFT supplemented by the scalar $S$ are only valid in the case in which all other NP degrees of freedom are much heavier than the resonance itself - an assumption that is not automatically justified. Furthermore, even if the assumption is valid, the ``SMEFT+$S$'' can only separate the scale $m_S$ from the heavier NP-scale and not $m_S$ from $\mu_\mathrm{SM}$. Therefore, in either case of the NP scale hierarchy, the result has to be matched onto the SCET$_\mathrm{BSM}$ to resum the large double logarithms.

The size of the resummation effects on the decay rates of the resonance were found to be large - ranging from $\sim 20\%$ in the mildest case up to a suppression factor of $\sim 85\%$ for the most extreme scenario and assuming a resonance mass of $m_S\approx 2.5$~TeV. The impact on predictions for and constraints on beyond-the-SM constructions is therefore significant and should be taken into account. Our framework provides a straightforward way of doing so, albeit at the time of writing only for a spin-0, gauge-singlet resonance. In future work, we will extend the approach to more complicated cases of non-singlet resonances and ones with non-zero spin, to cover the interesting cases like leptoquarks, $Z'$ bosons and heavy gluon excitations.

\section*{Acknowledgments}
MK gratefully acknowledges support by the Swiss National Science Foundation (SNF) under contract 200021-159720 and would like to thank the organizers for keeping up the spirit of Moriond by organizing a fantastic conference.

\end{document}